\RequirePackage{lineno}
\documentclass[twocolumn,10pt,prl,superscriptaddress,showpacs,
amssymb,amsmath,amsfonts,nofootinbib,floatfix]{revtex4}

\usepackage{graphicx} % Include figure files
\usepackage{latexsym}
\usepackage{dcolumn} % Align table columns on decimal point
\usepackage{bm} % bold math

\usepackage{epsfig}

\def\ba{\begin{eqnarray}}\def\ea{\end{eqnarray}}
\def\bc{\begin{center}}\def\ec{\end{center}}

\begin{document}
\pagestyle{empty}
%\linenumbers
\preprint{}
\title{  High Precision Measurement of Compton Scattering in the 5 GeV region  }

\author{P.~Ambrozewicz\footnote{Corresponding author}}
\affiliation{North Carolina A\&T State University, Greensboro, NC 27411, USA}
\affiliation{{\rm{Currently at:}} Eastern Virginia Medical School, Norfolk, VA 23501, USA}

\author{L. Ye}
\affiliation{Mississippi State University, Mississippi State, MS 39762, USA}

\author{Y.~Prok}
\affiliation{Massachusetts Institute of Technology, Cambridge, MA 02139, USA}
\affiliation{{\rm{Currently at:}} Virginia Commonwealth University, Richmond, VA 23284, USA}

\author{I.~Larin}
\affiliation{Alikhanov Institute for Theoretical and Experimental Physics, NRC "Kurchatov Institute", Moscow 117218, Russia}
\affiliation{University of Massachusetts, Amherst, MA 01003, USA}

\author{A.~Ahmidouch}
\affiliation{North Carolina A\&T State University, Greensboro, NC 27411, 
USA}

%\author{A.~Asratyan}
%\affiliation{Alikhanov Institute for Theoretical and Experimental Physics, Moscow, 
%Russia}

\author{K.~Baker}
\affiliation{Hampton university, Hampton, VA 23606, USA}

\author{V.~Baturin}
\affiliation{Thomas Jefferson National Accelerator Facility, Newport News, VA 23606,
USA}

\author{L.~Benton}
\affiliation{North Carolina A\&T State University, Greensboro, NC 27411,
USA}

\author{A.~Bernstein}
\affiliation{Massachusetts Institute of Technology, Cambridge, MA 02139, USA}

\author{V.~Burkert}
\affiliation{Thomas Jefferson National Accelerator Facility, Newport News, VA 23606,
USA}

\author{E.~Clinton}
\affiliation{University of Massachusetts, Amherst, MA 01003, USA}

\author{P.L.~Cole}
\affiliation{Idaho State University, Pocatello, ID 83209, USA}
\affiliation{{\rm{Currently at:}} Lamar University, Beaumont, TX 77710, USA}

\author{P.~Collins}
\affiliation{Arizona State University, Tempe, AZ 85281, USA}

\author{D.~Dale\footnote{Spokesperson}}
\affiliation{Idaho State University, Pocatello, ID 83209, USA}

\author{S.~Danagoulian$^{\dagger}$}
\affiliation{North Carolina A\&T State University, Greensboro, NC 27411,
USA}

\author{G.~Davidenko}
\affiliation{Alikhanov Institute for Theoretical and Experimental Physics, NRC "Kurchatov Institute", Moscow 117218, Russia}

\author{R.~Demirchyan}
\affiliation{North Carolina A\&T State University, Greensboro, NC 27411,
USA}

\author{A.~Deur}
\affiliation{Thomas Jefferson National Accelerator Facility, Newport News, VA 23606,
USA}

\author{A.~Dolgolenko}
\affiliation{Alikhanov Institute for Theoretical and Experimental Physics, NRC "Kurchatov Institute", Moscow 117218, Russia}

\author{D. Dutta}
\affiliation{Mississippi State University, Mississippi State, MS 39762,
USA}

\author{G.~Dzyubenko}
\affiliation{Alikhanov Institute for Theoretical and Experimental Physics, NRC "Kurchatov Institute", Moscow 117218, Russia}

%\author{R.~Ent}
%\affiliation{Thomas Jefferson National Accelerator Facility, Newport News, VA 23606,
%USA}

\author{A.~Evdokimov}
\affiliation{Alikhanov Institute for Theoretical and Experimental Physics, NRC "Kurchatov Institute", Moscow 117218, Russia}
\affiliation{{\rm{Currently at:}} University of Illinois at Chicago, Chicago, USA}

\author{G.~Fedotov}
\affiliation{Moscow State University, Moscow 119991, Russia}
\affiliation{{\rm{Currently at:}} National Research Centre ”Kurchatov Institute” B. P.
Konstantinov Petersburg Nuclear Physics Institute, Gatchina, St. Petersburg, Russia}

\author{J.~Feng}
\affiliation{University of North Carolina Wilmington, Wilmington, NC 28403, USA}

\author{M.~Gabrielyan}
\affiliation{University of Kentucky, Lexington, KY 40506, USA}

\author{L.~Gan$^{\dagger}$}
\affiliation{University of North Carolina Wilmington, Wilmington, NC 28403, USA}

\author{H. Gao}
\affiliation{Duke University and Triangle University Nuclear Lab, Durham, MC 27708, USA}

\author{A.~Gasparian\footnote{Spokesperson, contact person}}
\affiliation{North Carolina A\&T State University, Greensboro, NC 27411, USA}

\author{N.~Gevorkyan}
\affiliation{Yerevan Physics Institute, Yerevan 0036, Armenia}

\author{S.~Gevorkyan}
\affiliation{Joint Institute for Nuclear Research, Dubna 141980, Russia, 
On leave of absence from Yerevan Physics Institute, Yerevan, Armenia}

\author{A.~Glamazdin}
\affiliation{Kharkov Institute of Physics and Technology, Kharkov 310108, Ukraine}

\author{V.~Goryachev}
\affiliation{Alikhanov Institute for Theoretical and Experimental Physics, NRC "Kurchatov Institute", Moscow 117218, Russia}

\author{L.~Guo}
\affiliation{Florida International University, Miami, FL 33199, USA}

\author{V.~Gyurjyan}
\affiliation{Thomas Jefferson National Accelerator Facility, Newport News, VA 23606,
USA}

\author{K.~Hardy}
\affiliation{North Carolina A\&T State University, Greensboro, NC 27411, USA}

\author{J.~He}
\affiliation{Institute of High Energy Physics, Chinese Academy of Sciences,
Beijing 100049, China}

\author{E.~Isupov}
\affiliation{Moscow State University, Moscow 119991, Russia}

\author{M.M.~Ito}
\affiliation{Thomas Jefferson National Accelerator Facility, Newport News, VA 23606,
USA}

\author{L.~Jiang}
\affiliation{University of North Carolina Wilmington, Wilmington, NC 28403, USA}

\author{H.~Kang}
\affiliation{Seoul National University, Seoul 08826, Korea}

\author{D.~Kashy}
\affiliation{Thomas Jefferson National Accelerator Facility, Newport News, VA 23606,
USA}

\author{M.~Khandaker$^{\dagger}$}
\affiliation{Norfolk State University, Norfolk, VA 23504, USA}

\author{P.~Kingsberry}
\affiliation{Norfolk State University, Norfolk, VA 23504, USA}

\author{F.~Klein}
\affiliation{The Catholic University of America, Washington, DC 20064, USA}

\author{A.~Kolarkar}
\affiliation{University of Kentucky, Lexington, KY 40506, USA}

\author{M. Konchatnyi}
\affiliation{Kharkov Institute of Physics and Technology, Kharkov 310108, Ukraine}

\author{O. Korchin}
\affiliation{Kharkov Institute of Physics and Technology, Kharkov 310108,  Ukraine}

\author{W. Korsch}
\affiliation{University of Kentucky, Lexington, KY 40506, USA}

\author{O.~Kosinov}
\affiliation{Idaho State University, Pocatello, ID 83209, USA}

\author{S.~Kowalski}
\affiliation{Massachusetts Institute of Technology, Cambridge, MA 02139, USA}

\author{M.~Kubantsev}
\affiliation{Northwestern University, Evanston, IL 60208, USA}

\author{A. Kubarovsky}
\affiliation{Thomas Jefferson National Accelerator Facility, Newport News, VA 23606,
USA}

\author{V. Kubarovsky}
\affiliation{Thomas Jefferson National Accelerator Facility, Newport News, VA 23606,
USA}

\author{D.~Lawrence}
\affiliation{Thomas Jefferson National Accelerator Facility, Newport News, VA 23606,
USA}

\author{X.~Li}
\affiliation{University of North Carolina Wilmington, Wilmington, NC 28403, USA}

\author{M.~Levillain}
\affiliation{North Carolina A\&T State University, Greensboro, NC 27411,
USA}

\author{H.~Lu}
\affiliation{Carnegie Mellon University, Pittsburgh, PA 15213, USA}

\author{L.~Ma}
\affiliation{University of North Carolina Wilmington, Wilmington, NC 28403, USA}

\author{P.~Martel}
\affiliation{University of Massachusetts, Amherst, MA 01003, USA}

\author{V.~Matveev}
\affiliation{Alikhanov Institute for Theoretical and Experimental Physics, NRC "Kurchatov Institute", Moscow 117218, Russia}

\author{D.~McNulty}
\affiliation{Massachusetts Institute of Technology, Cambridge, MA 02139, USA}
\affiliation{{\rm{Currently at:}} Idaho State University, Pocatello, ID 83209, USA}

\author{B.~Mecking}
\affiliation{Thomas Jefferson National Accelerator Facility, Newport News, VA 23606,
USA}

\author{A.~Micherdzinska}
\affiliation{George Washington University, Washington, DC 20064, USA}

\author{B.~Milbrath}
\affiliation{Pacific Northwest National Laboratory, Richland, WA 99354, USA}

\author{R.~Minehart}
\affiliation{University of Virginia, Charlottesville, VA 22094, USA}

\author{R.~Miskimen$^{\dagger}$}
\affiliation{University of Massachusetts, Amherst, MA 01003, USA}

\author{V.~Mochalov}
\affiliation{Institute for High Energy Physics, Protvino 142280, Russia}

\author{B.~Morrison}
\affiliation{Arizona State University, Tempe, AZ 85281, USA}

\author{S.~Mtingwa}
\affiliation{North Carolina A\&T State University, Greensboro, NC 27411, USA}

\author{I.~Nakagawa}
\affiliation{University of Kentucky, Lexington, KY 40506, USA}

\author{S.~Overby}
\affiliation{North Carolina A\&T State University, Greensboro, NC 27411, USA}

\author{E.~Pasyuk}
\affiliation{Arizona State University, Tempe, AZ 85281, USA}
\affiliation{{\rm{Currently at:}} Thomas Jefferson National Accelerator Facility, Newport News, VA 23606, USA}

\author{M.~Payen}
\affiliation{North Carolina A\&T State University, Greensboro, NC 27411, USA}

\author{K.~Park}
\affiliation{Thomas Jefferson National Accelerator Facility, Newport News, VA 23606,
USA}

\author{R~Pedroni}
\affiliation{North Carolina A\&T State University, Greensboro, NC 27411, USA}

\author{W.~Phelps}
\affiliation{Christopher Newport University, Newport News, VA 23606, USA}

\author{D.~Protopopescu}
\affiliation{Glasgow University, Glasgow G12 8QQ, UK}

\author{D.~Rimal}
\affiliation{Florida International University, Miami, FL 33199, USA}

\author{B.G.~Ritchie}
\affiliation{Arizona State University, Tempe, AZ 85281, USA}

\author{D.~Romanov}
\affiliation{Thomas Jefferson National Accelerator Facility, Newport News, VA 23606,
USA}

\author{C~Salgado}
\affiliation{Norfolk State University, Norfolk, VA 23504, USA}

\author{A.~Shahinyan}
\affiliation{Yerevan Physics Institute, Yerevan 0036, Armenia}

\author{A~Sitnikov}
\affiliation{Alikhanov Institute for Theoretical and Experimental Physics, NRC "Kurchatov Institute", Moscow 117218, Russia}

\author{D.~Sober}
\affiliation{The Catholic University of America, Washington, DC 20064, USA}

\author{S.~Stepanyan}
\affiliation{Thomas Jefferson National Accelerator Facility, Newport News, VA 23606,
USA}

\author{W.~Stephens}
\affiliation{University of Virginia, Charlottesville, VA 22094, USA}

\author{V.~Tarasov}
\affiliation{Alikhanov Institute for Theoretical and Experimental Physics, NRC "Kurchatov Institute", Moscow 117218, Russia}

\author{S.~Taylor}
\affiliation{Thomas Jefferson National Accelerator Facility, Newport News, VA 23606,
USA}

\author{A.~Teymurazyan}
\affiliation{University of Kentucky, Lexington, KY 40506, USA}

\author{J.~Underwood}
\affiliation{North Carolina A\&T State University, Greensboro, NC 27411, USA}

\author{A.~Vasiliev}
\affiliation{Institute for High Energy Physics, Protvino 142280, Russia}

%\author{V.~Verebryusov}
%\affiliation{Alikhanov Institute for Theoretical and Experimental Physics, Moscow,
%Russia}

\author{V.~Vishnyakov}
\affiliation{Alikhanov Institute for Theoretical and Experimental Physics, NRC "Kurchatov Institute", Moscow 117218, Russia}

\author{D.~P.~Weygand}
\affiliation{Thomas Jefferson National Accelerator Facility, Newport News, VA 23606,
USA}
\author{M.~Wood}
\affiliation{University of Massachusetts, Amherst, MA 01003, USA}

\author{Y.~Zhang}
\affiliation{Duke University and Triangle University Nuclear Lab, Durham, MC 27708, USA}

\author{S.~Zhou}
\affiliation{Chinese Institute of Atomic Energy, Beijing 102413, China}

\author{B.~Zihlmann}
\affiliation{Thomas Jefferson National Accelerator Facility, Newport News, VA 23606,
USA}
\collaboration{The {\it PrimEx} Collaboration}

\date{\today} % It is always \today, but any date may be explicitly specified 

\begin{abstract}
\vspace{9ex}
The cross section of atomic electron Compton scattering $\gamma + e \rightarrow \gamma^\prime + e^\prime $ 
was measured in the 4.400--5.475 GeV photon beam energy region by the {\em PrimEx} collaboration 
at Jefferson Lab with an accuracy of 2.6\% and less. The results are consistent with theoretical 
predictions that include next-to-leading order radiative corrections. The measurements provide the 
first high precision 
test of this elementary QED process at beam energies greater than 0.1 GeV.

\end{abstract}

 % PACS

\maketitle
\section{I. Introduction}
Quantum electrodynamics (QED) is one of  the most successful theories in modern physics;
and the Compton scattering of photons by free electrons 
$\gamma + e \rightarrow \gamma^\prime + e^\prime$
is the simplest and the most elementary pure QED process. The lowest-order Compton scattering
diagrams (see Fig.~\ref{comp-cross1}) were first calculated by Klein
and Nishina in 1929 \cite{Klein29} and by Tamm in 1930 \cite{Tamm30}. Higher-order contributions arising 
from the interference between the leading order single Compton scattering amplitude and the radiative and 
double Compton scattering amplitudes were calculated in the 1950s~\cite{radiative1,Mandl52}. 
Figure~\ref{comp-cross1} shows the Feynman diagrams illustrating these two processes. They were 
subsequently re-evaluated in the 60s and early 70s to make them amenable for calculation using 
modern computational techniques~\cite{radiative2}-\cite{Ram71}. Corrections to the 
leading order Compton total cross section at the level of a few percent are predicted for beam energies
above 0.1 GeV~\cite{radiative3}, hence the next-to-leading order (NLO) corrections are important 
when studying Compton scattering at these energies. \\

\indent Experiments performed so far were mostly in the energy region
below 0.1 GeV; a few experiments probed the 0.1-1.0 GeV energy range with a precision of
10--15\%~\cite{Coensgen53}-\cite{Gittlman68}.  
Only one experiment \cite{Goshaw78} measured the Compton scattering total cross section 
up to 5.0 GeV using a bubble-chamber detection technique. The experimental uncertainties for
energies above 1 GeV were at the level of 20--70\%.
Due to the lack of precise data, higher order corrections to the Klein-Nishina
formula have never been tested experimentally. 
This paper reports on new measurements of the Compton scattering cross section with a precision 
of 1.7\% performed by the {\em PrimEx} collaboration at Jefferson Lab (JLab) for two separate 
running periods. 
The total  cross sections  in a forward direction
on $^{12}$C and $^{28}$Si targets were measured in the 4.400--5.475 GeV-energy region.
The  precision achieved by this experiment provides, for the first time, an important 
test of the QED prediction for the Compton scattering process with corrections to the order of
$\mathcal{O}$($\alpha$), where $\alpha$ is the fine structure constant. In this article, we will 
summarize the theoretical calculations (Sec. II), describe our experimental procedure (Sec. III), and 
present the results of the comparison between the data and the theoretical predictions (Sec. IV).

\section{II. A summary of theoretical calculations}
\begin{figure}[htb!]
\begin{center}
\epsfig{angle=0,width=8.cm,file=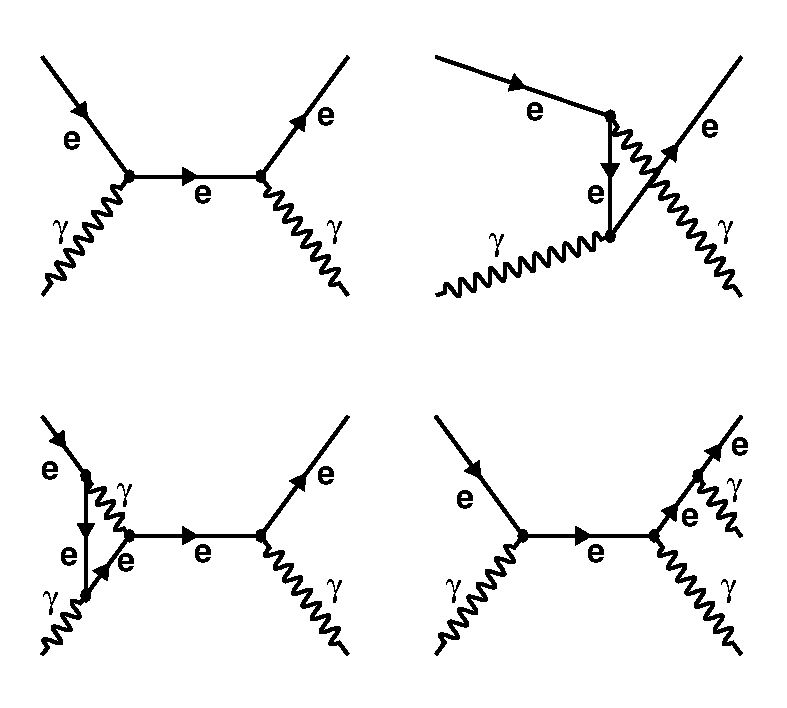}
\end{center} 
\caption{(Top) The lowest-order Feynman diagrams for single Compton scattering. (Bottom) Typical radiative correction (Left), and 
double Compton scattering contributions (Right) to single Compton scattering.}
\label{comp-cross1}
\end{figure}

The leading order Compton scattering cross section (see Fig.~\ref{comp-cross1}, top) was first
calculated  by Klein and Nishina~\cite{Klein29} and the result is known as the Klein-Nishina 
formula~\cite{Peskin95}:

\begin{equation}
\frac{d\sigma}{d\Omega}=\frac{r_e^2}{2}\frac{1}{[1+\gamma (1-\cos\theta)]^2}
 [1+\cos^2\theta+\frac{\gamma^2 (1-\cos\theta)^2}{1+\gamma (1-\cos\theta)}]
 \nonumber
\end{equation}

\noindent where $r_e$ is the classical electron radius, $\gamma$ is the ratio of the photon 
beam energy to the mass 
of electron, and $\theta$ is the photon scattering angle. 
This formula predicts that the Compton scattering at high energies has two basic features: 
(i) the total cross section decreases with increasing beam energy, 
$E$, as approximately $1/E$, and (ii) the differential cross section is sharply peaked 
at small angles relative to the incident photons. 

The theoretical foundation for the next-to-leading order radiative corrections to the 
Klein-Nishina formula had been well established by early 70s. The radiative corrections to 
$\mathcal{O}$($\alpha$) were initially evaluated by Brown and Feynman~\cite{radiative1} 
in 1952. This correction is caused by two types of processes. The first type, a 
virtual-photon correction, arises from the possibility that the electron may emit and 
reabsorb a virtual photon in the scattering process (see bottom left panel of Fig.~\ref{comp-cross1}). 
The second type is a soft-photon double Compton effect, in which the energy of one of
the emitted photons is much smaller than the electron mass ($\omega_2<\omega_{2max}\ll m_e$, 
where $\omega_2$ is the energy of the additional photon, $\omega_{2max}$ is a cut-off energy, and
$m_e$ is the electron mass),  as shown in the bottom right panel of
Fig.~\ref{comp-cross1}. These two contributions must be  taken into account together 
since it is impossible to separate them experimentally. Moreover, the infrared divergence 
term from the virtual-photon process is canceled by the infrared divergence term in the 
soft-photon double Compton process, resulting in a finite physically meaningful correction
($\delta^{SV}$). The value of $\delta^{SV}$, where $SV$ stands for $S$(oft) and 
$V$(irtual), is predicted to be negative as described by Eq.~(2.6) and Eq.~(2.15) 
in~\cite{radiative3}.\\

On the other hand, a hard-photon double Compton effect occurs when 
both emitted photons in the double Compton process have energies larger than the cut-off 
energy, $\omega_{2max}$. When comparing the experimental result with the theoretical 
calculation, one must also take into account the contributions from the hard-photon double
Compton effect since the experimental apparatus has finite resolutions leading to
limitations on the measurements of both energies and angles~\cite{radiative3}.  The 
differential cross section of the double Compton effect was initially calculated by Mandl and 
Skyrme~\cite{Mandl52}, and  the total cross section of the Hard-photon Double Compton 
process ($\delta^{HD}$) is described by Eq.~(6.6) in reference~\cite{radiative3}
and its value is 
predicted to be positive.  Summing up $\delta^{SV}$ and $\delta^{HD}$, the total NLO correction to the 
total cross section is predicted to be a few percent for photon beam energies  up to 10 GeV.\\

In order to interpret the experimental results and compare with the theoretical 
predictions,  one needs to develop a reliable numerical method to integrate the cross 
section and calculate the radiative corrections incorporating the experimental resolutions.
The latter is critical in calculating the contribution from the hard photon double Compton 
effect correctly. As discussed above, the corrections are divided into two types 
($\delta^{SV}$ and $\delta^{HD}$) depending on whether the energy of the secondary emitted
photon is less or greater than an arbitrary energy scale, denoted by  $\omega_{2max}$,
which should be much smaller than the electron mass~\cite{radiative3}. Since the 
physically measurable cross section contains the corrections from both types, the final 
integrated total cross section must be independent of the values of $\omega_{2max}$. Two 
different methods had been developed to prove this independence. \\

The first method~\cite{PrimEx42} is based on the BASES/SPRING Monte Carlo simulation 
package~\cite{Kawabata}. BASES uses the stratified sampling method to integrate the 
differential cross section, and SPRING uses the probability information obtained 
during the BASES integration to generate Compton events. The parameter $\omega_{2max}$ does not enter 
the differential cross section 
explicitly but is contained in the limits of integration over the energy. For a consistency
check, the total cross section was calculated with several values of $\omega_{2max}$. While
the calculated total Klein-Nishina cross section corrected with the virtual and soft photon
processes ($\sigma_{SV}$) as well as the total hard photon double Compton cross section 
($\sigma_{HD}$), both, depend on the $\omega_{2max}$ parameter, the sum of the two 
corrections ($\sigma_{SV}+\sigma_{HD}$) is independent, within 0.1\%, of the choice of 
$\omega_{2max}$, as expected. \\

The second numerical method was developed by M.~Konchatnyi~\cite{PrimEx37}, where the 
parameter $\omega_{2max}$ is analytically removed from the integration. The total Compton scattering cross
section on ${}^{12}$C along with radiative corrections, calculated using 
the two numerical methods~\cite{PrimEx42}~\cite{PrimEx37} described above, were compared to each other and to the XCOM~\cite{Berger} database of the 
National Institute of Standards and Technology ({\slshape NIST}). The cross sections were found to be consistent with each other. The radiative corrections are about 4\% of the total cross section for a beam energy of $\sim$5~GeV. In the data analysis described below, the BASES/SPRING method is used
to calculate the radiatively corrected cross section and to generate events for the experimental acceptance study.

\section{III. Experimental Procedure}

\begin{figure}[htb!]
\centering
\epsfig{file=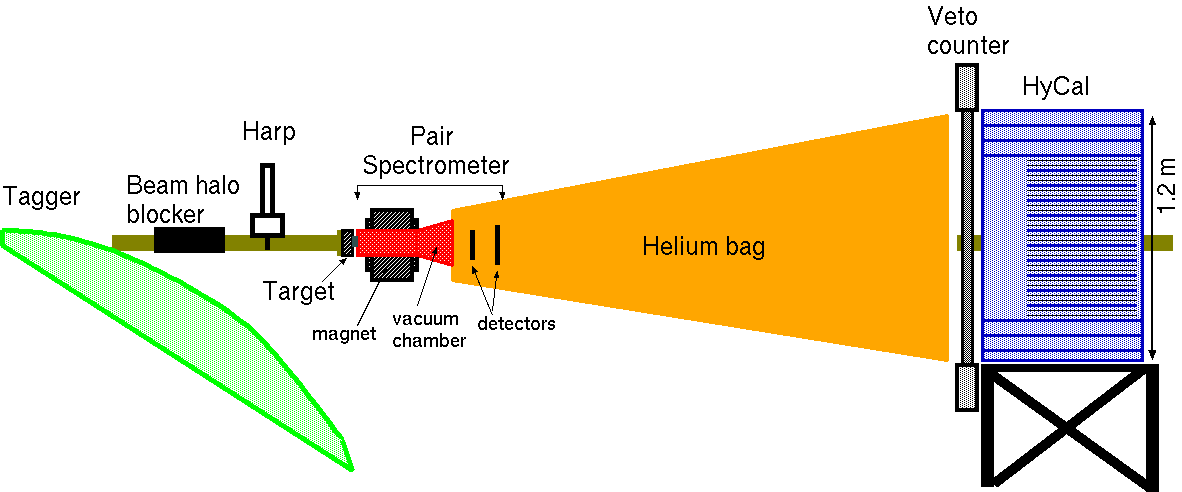,width=8.5cm}
\caption{ Diagram, not to scale, of the experimental setup. The pair spectrometer placed between the target and the helium bag, had the magnet turned off during the Compton experiment.}
\label{figLayout}
\end{figure}

The atomic electron Compton scattering process $\gamma + e \rightarrow \gamma^{'} + e'$ was 
measured using the apparatus built for the {\em PrimEx} experiment~\cite{ConceptualD}, 
which aimed to measure the $\pi^0$ lifetime and was performed over two run periods in 2004 and 2010, 
in Hall~B at JLab. The Compton scattering data were collected periodically, once  per week 
during both running periods. The primary experimental equipment included (see Fig.~\ref{figLayout}): 
(i)~the existing Hall~B high intensity and high resolution photon tagger~\cite{Sober}, which 
provides the timing and energy information 
of incident photons up to 6 GeV; (ii) solid production targets~\cite{Martel}: 
$^{12}$C (5\% radiation length (r.l.)), used during the first running period, and $^{12}$C (8\% 
r.l.) and $^{28}$Si (10\% r.l.) added in the second running period; 
(iii)~a pair spectrometer (PS), located downstream
of the production target, to continuously measure the relative photon tagging ratio~\cite{Teymurazyan}, and 
consequently the absolute photon flux, which was obtained by normalizing to the absolute photon 
tagging efficiency measured periodically with a total absorption counter (TAC) at low beam 
intensities (not shown in Fig.~\ref{figLayout}); 
(iv)~a $118\times118$~cm$^2$ high resolution hybrid calorimeter (HyCal~\cite{Gasparian}) with
12 scintillator charge particle veto counters, which were located $\sim$7~m 
downstream of the target, to detect forward scattered electromagnetic particles; and (v)~a scintillator 
fiber based photon beam profile and position detector located behind HyCal for online beam position 
monitoring (not shown in Fig.~\ref{figLayout}). \\

To minimize the photon conversion and electron 
multiple scattering, the gap between the PS magnet and the HyCal was occupied by a 
plastic foil container filled with helium at atmospheric pressure.
The energies and positions of the scattered photon and electron were measured by the  HyCal 
calorimeter. In conjunction with the beam energy (4.9-5.5 GeV during the first experiment and
4.4-5.3 GeV during the second one), which was measured by the photon tagger, the complete kinematics
of the Compton events was determined. During the Compton runs the experimental setup was 
identical to the one 
used for the $\pi^0$ production runs, except for the pair spectrometer magnet being turned off 
to allow detection of both scattered photons and recoiling electrons in the calorimeter. 
The use of the same experimental apparatus, as well as the similar kinematics allowed 
the measurement of the Compton cross section to be employed as a tool to verify the systematic uncertainty 
of the $\pi^0$ experiments. The photon flux, target thickness and HyCal energy response being some of the common systematic uncertainty between the Compton and $\pi^0$ run types and resulting analyses.\\

A coincidence between the photon tagger in the energy interval of % (2.51--2.63~GeV or 
4.4--5.5~GeV and the HyCal calorimeter with a total energy deposition greater than 2.5 GeV 
formed an event trigger for the first running period, while in the second running period the 
trigger was defined just by the total energy deposition greater than 2.5 GeV in the HyCal. 
The event selection criteria were: 
(i) the time difference between the incident photon, $\rm{t_{Tag}}$
and the scattered particles detected by the HyCal calorimeter, 
$\rm{t_{HyCal}}$ had to be $|\rm{t_{Tag}-t_{HyCal}}| <5\sigma_{t} $, where $\sigma_{t}$ =1.03 ns 
is the timing resolution of the detector system.
(ii) the difference in the azimuthal angle between the 
scattered photon and electron had to be $|\Delta\phi| <5 \sigma_{\phi}$, where $\sigma_{\phi} =$7$^{\circ}$
is the azimuthal angular resolution for the first running period, 
(for the second running period a target dependent
resolution of $\sigma_{\phi} = $4.0 - 4.7$^{\circ}$ was used); 
\noindent (iii) the reconstructed reaction vertex 
position was required to be consistent with the target thickness and position; 
(iv) the spatial distance between the scattered photon and electron as detected by the HyCal calorimeter
had to be larger than a photon energy dependent minimum separation resulting from the reaction 
being elastic; 
the minimum separation of 16~cm for the first running period and
$R_{min}(E) = $19.0 - 1.95$\times$(4.85-$E$) for the second running period; 
and (v) the difference between the incident photon 
energy as measured by the tagger, $\rm{E_{Tag}}$ and the reconstructed incident photon energy, 
$\rm{E_{HyCal}}$, had to be $|\rm{E_{Tag}-E_{HyCal}}| <$1 (0.4)~GeV for the first (second) running
period. In the event reconstruction, the measured energy of the more energetic scattered  particles
(photon or electron) and the coordinate information of both scattered particles detected by the 
calorimeter was used. The offline energy detection threshold per particle in the HyCal calorimeter
was 0.5~GeV.

\begin{figure}[htb]
\centering
\epsfig{file=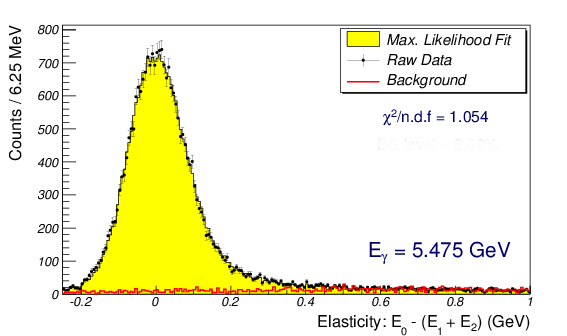,width=9.cm}
\caption{An example of the fit to the elasticity distribution for the highest energy bin. The background contribution is shown in red.  $E_0$, $E_1$ + $E_2$ are the beam and sum of scattered electron and scattered $\gamma$ energies respectively. Note that since particle identification was not used, the photons and electrons cannot be distinguished from each other. The cluster with higher energy is designated as $E_1$ and the cluster with lower energy is designated as $E_2$.}\label{yfit-tc01}
\end{figure}

To extract the Compton yields, the signal and background events (at a level of several percent
of the yield) were separated for every incident photon energy bin (with a width of $\sim$1\% of 
the nominal beam energy). 
The background originating from the target ladder and housing was determined using data from dedicated 
empty target runs, and the yields from these runs were normalized to the beam current and subtracted away.
The remaining events that passed all of the five selection criteria described above were used to form an 
elasticity distribution, $\Delta E = E_0 - (E_1 + E_2)$ where $E_0$ is the measured energy of the incident photon, and $E_1 + E_2 = E_{\gamma^\prime} + E_{e^\prime}$ is the sum of the scattered photon ($E_{\gamma^\prime}$) and the scattered electron ($E_{e^\prime}$) energies, which were either measured (the first experiment) or calculated using the Compton scattering kinematics (the second experiment). The cluster with higher energy is designated as $E_1$ and the cluster with lower energy is designated as $E_2$.
The elasticity distribution was then fit to the simulated signal and background distributions, 
using a maximum likelihood method~\cite{Barlow}. Their overall amplitudes were parameters in the fit, as 
shown in Fig. \ref{yfit-tc01}, which is a typical result. The reduced $\chi^2$ for the different energy bins and for the 3 targets varied between 0.9 - 1.7. \\

The signal was generated by a Monte Carlo simulation employing the BASES/SPRING package as described in 
Sec. II~\cite{PrimEx42},\cite{Kawabata},  which  included the 
radiative processes and the double Compton contribution. The simulated signal events were propagated 
through a {\slshape GEANT}-based simulation of the experimental apparatus and then processed using the same event reconstruction software that was used to extract the experimental yield. The {\slshape GEANT} based simulation framework includes parameters such as: light yield and transparency of PbWO$_4$ and lead-glass modules, the non-uniformity of the light yield (which was adjusted to match the light yield measured at IHEP, Protvino, Russia~\cite{ilya}), and the quantum efficiency of the photo-multiplier tubes (Hamamatsu-R4125HA). The Monte Carlo parameters were tuned to reproduce the measured resolutions (position and energy) to within 3\%, primarily for the calibration data obtained from the scan of the calorimeter with a tagged photon beam. The calibration scans were simulated to have the same statistics in the energy and position spectra as in the data, such that the uncertainty in the calibration constants for the data and simulation were the same. The energy response function and angular resolution was further verified by comparing the simulation to  dedicated data collected using a thin 0.5\% r.l. Be target and was found to be consistent within 3\%~\cite{Gasparian2}. \\

The shape of the background was modeled by the accidental events alone for the first running period, while the pair production channel 
was also included for the second running period. The accidental background was selected from the data using the events that were outside the coincidence time window, from  $|\rm{t_{Tag}-t_{HyCal}}| >5\sigma_{t}$, 
but satisfied the remaining four criteria described above. The pair production contribution was generated 
using the {\slshape GEANT} simulation 
toolkit with its results handled in the same manner as the experimental yield. 
The amplitude from the maximum likelihood fit was then used to subtract the background from 
the experimental yield for each incident photon energy bin, giving the Compton yield.

\section{IV. Results}

\begin{figure}[htb!]
\hspace*{-4mm}
\epsfig{file=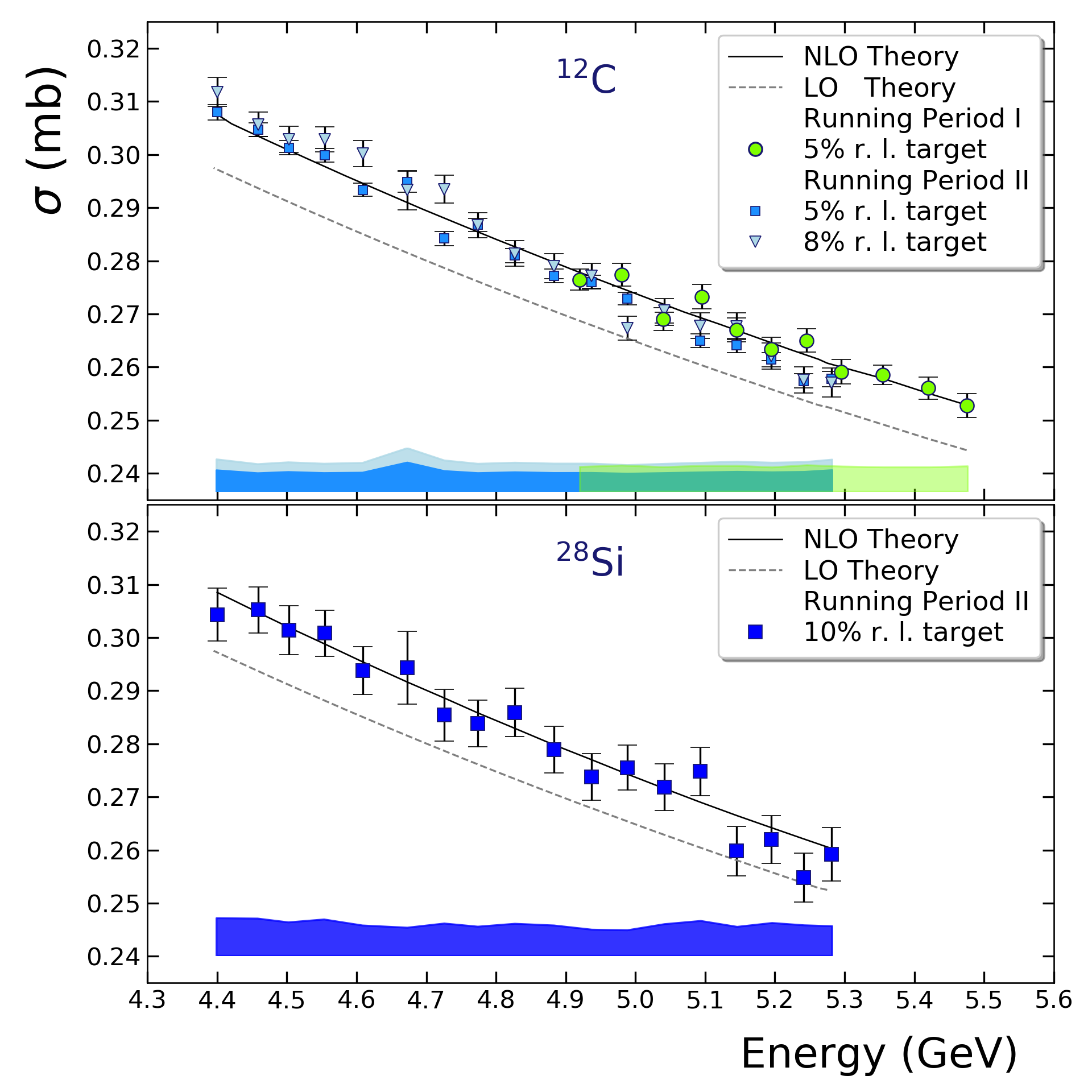,width=9cm\vspace*{-2mm}}
\caption{The Compton cross sections measured on atomic electrons 
of $^{12}$C and $^{28}$Si targets. 
The dashed curve corresponds to the Klein-Nishina calculation. The solid curve is 
the result of next-to-leading order calculation.
Error bars are statistical uncertainties. Bands at the bottom of both plots show 
point-to-point systematic uncertainties.
}\label{csec3}
\end{figure}

\begin{table}[htbp!]
\footnotesize
\begin{tabular}{|l|c|c|c|c|}\hline
& \multicolumn{4}{c}{\bf{Running Period}} \vline \\ \cline{2-5}
&  & \multicolumn{3}{c}{} \vline \\
& \bf{I} & \multicolumn{3}{c}{\bf{II}} \vline \\  \hline %\cline{2-5}
& & & & \\
 \bf{Source Of Uncertainty} & ${}^{\mathbf{12}}$\bf{C} & ${}^{\mathbf{12}}$\bf{C}(5\%) & ${}^{\mathbf{12}}$\bf{C}(8\%) &${}^{\mathbf{28}}$\bf{Si} \\  \hline \hline
photon flux & 1.0  & 0.82 & 0.82 & 0.82 \\  \hline 
target composition, thickness & 0.05 & 0.05 & 0.11 & 0.35 \\ \hline 
coincidence timing  & 0.05 & 0.49 & 0.68 & 0.60 \\ \hline
coplanarity  & 0.08  & 0.51 & 0.66  & 0.71\\ \hline
geometrical acceptance  & 0.60  & 0.29 & 0.31 & 0.62\\ \hline
background subtraction  & 0.72  & 1.07 & 1.32 & 1.31\\  \hline
HyCal energy response   & 0.5  & 0.5 & 0.5 & 0.5\\ \hline \hline 
total  & 1.5  & 1.6 &  1.9 & 2.0\\ \hline

\end{tabular}
\caption{Estimated systematic uncertainties, in percent, for each target over 
the entire energy range.}
\label{table-error}
\end{table}

\begin{table*}[ht!]
\begin{tabular}{c|c|c|c|c|c|c|c|c|c}
&  \multicolumn{3}{c|}{\bf{${\mathbf{{}^{12}}}$C - 5\% r. l.}} & \multicolumn{3}{c|}{\bf{${\mathbf{{}^{12}}}$C - 8\% r. l.}} & \multicolumn{3}{c}{\bf{${\mathbf{{}^{28}}}$Si - 10\% r. l.}} \\ 
\hline 
Energy (GeV) &   $\sigma$  &  $\delta_{stat}$  &  $\delta_{syst}$ &  $\sigma$  &  $\delta_{stat}$  &  $\delta_{syst}$ &  $\sigma$  &  $\delta_{stat}$  &  $\delta_{syst}$ \\
\hline
 4.40  & 0.3080  &  0.0014  &  0.0040  &  0.3118  &  0.0027  &  0.0062  &  0.3043  &  0.0050  &  0.0068 \\
 4.46  & 0.3047  &  0.0012  &  0.0039  &  0.3057  &  0.0023  &  0.0065  &  0.3052  &  0.0043  &  0.0068 \\
 4.50  & 0.3013  &  0.0013  &  0.0037  &  0.3029  &  0.0025  &  0.0051  &  0.3014  &  0.0046  &  0.0061 \\
 4.55  & 0.2999  &  0.0013  &  0.0043  &  0.3029  &  0.0024  &  0.0068  &  0.3008  &  0.0044  &  0.0066 \\
 4.61  & 0.2934  &  0.0013  &  0.0036  &  0.3002  &  0.0024  &  0.0053  &  0.2938  &  0.0045  &  0.0055 \\
 4.67  & 0.2949  &  0.0020  &  0.0046  &  0.2933  &  0.0037  &  0.0058  &  0.2943  &  0.0069  &  0.0051 \\
 4.73  & 0.2842  &  0.0014  &  0.0045  &  0.2935  &  0.0026  &  0.0051  &  0.2854  &  0.0049  &  0.0059 \\
 4.77  & 0.2868  &  0.0013  &  0.0047  &  0.2867  &  0.0024  &  0.0077  &  0.2838  &  0.0044  &  0.0053 \\
 4.83  & 0.2810  &  0.0013  &  0.0043  &  0.2814  &  0.0024  &  0.0039  &  0.2859  &  0.0046  &  0.0058 \\
 4.88  & 0.2772  &  0.0013  &  0.0063  &  0.2790  &  0.0024  &  0.0047  &  0.2789  &  0.0044  &  0.0055 \\
 {\bf{{\slshape{4.92}}}}  &  {\bf{{\slshape{0.2765}}}}  &  {\bf{{\slshape{0.0020}}}}  &  {\bf{{\slshape{0.0042}}}} & & & & & & \\
 4.94  & 0.2760  &  0.0013  &  0.0059  &  0.2772  &  0.0024  &  0.0058  &  0.2738  &  0.0044  &  0.0048 \\
 {\bf{{\slshape{4.98}}}}  &  {\bf{{\slshape{0.2774}}}}  &  {\bf{{\slshape{0.0021}}}}  &  {\bf{{\slshape{0.0044}}}} & & & & & & \\
 4.99  & 0.2729  &  0.0012  &  0.0059  &  0.2673  &  0.0022  &  0.0045  &  0.2755  &  0.0042  &  0.0047 \\
 {\bf{{\slshape{5.04}}}}  &  {\bf{{\slshape{0.2691}}}}  &  {\bf{{\slshape{0.0021}}}}  &  {\bf{{\slshape{0.0041}}}} & & & & & & \\ 
 5.04  & 0.2691  &  0.0013  &  0.0033  &  0.2706  &  0.0023  &  0.0061  &  0.2718  &  0.0044  &  0.0058 \\
 {\bf{{\slshape{5.09}}}}  &  {\bf{{\slshape{0.2732}}}}  &  {\bf{{\slshape{0.0023}}}}  &  {\bf{{\slshape{0.0044}}}} & & & & & & \\ 
 5.09  & 0.2650  &  0.0013  &  0.0045  &  0.2678  &  0.0024  &  0.0037  &  0.2748  &  0.0046  &  0.0063 \\
 {\bf{{\slshape{5.15}}}}  &  {\bf{{\slshape{0.2670}}}}  &  {\bf{{\slshape{0.0023}}}}  &  {\bf{{\slshape{0.0043}}}} & & & & & & \\ 
 5.15  & 0.2641  &  0.0013  &  0.0047  &  0.2677  &  0.0025  &  0.0072  &  0.2598  &  0.0046  &  0.0053 \\
 {\bf{{\slshape{5.20}}}}  &  {\bf{{\slshape{0.2634}}}}  &  {\bf{{\slshape{0.0022}}}}  &  {\bf{{\slshape{0.0041}}}} & & & & & & \\  
 5.20  & 0.2614  &  0.0013  &  0.0035  &  0.2621  &  0.0024  &  0.0055  &  0.2620  &  0.0045  &  0.0060 \\
 5.24  & 0.2574  &  0.0013  &  0.0037  &  0.2576  &  0.0025  &  0.0054  &  0.2548  &  0.0046  &  0.0056 \\
 {\bf{{\slshape{5.25}}}}  &  {\bf{{\slshape{0.2650}}}}  &  {\bf{{\slshape{0.0022}}}}  &  {\bf{{\slshape{0.0045}}}} & & & & & & \\
 5.28  & 0.2578  &  0.0014  &  0.0063  &  0.2571  &  0.0027  &  0.0040  &  0.2592  &  0.0050  &  0.0054 \\
 {\bf{{\slshape{5.29}}}}  &  {\bf{{\slshape{0.2591}}}}  &  {\bf{{\slshape{0.0023}}}}  &  {\bf{{\slshape{0.0042}}}} & & & & & &  \\
 {\bf{{\slshape{5.36}}}}  &  {\bf{{\slshape{0.2585}}}}  &  {\bf{{\slshape{0.0018}}}}  &  {\bf{{\slshape{0.0041}}}} & & & & & &  \\
 {\bf{{\slshape{5.42}}}}  &  {\bf{{\slshape{0.2560}}}}  &  {\bf{{\slshape{0.0021}}}}  &  {\bf{{\slshape{0.0041}}}} & & & & & &  \\
 {\bf{{\slshape{5.47}}}}  &  {\bf{{\slshape{0.2528}}}}  &  {\bf{{\slshape{0.0022}}}}  &  {\bf{{\slshape{0.0043}}}} & & & & & &  \\
\hline 
\end{tabular}
\caption{Compton cross sections. Bold italicized numbers correspond to values obtained during the first 
running period.}
\label{table-csecs}
\end{table*}

The Compton scattering total cross sections (TABLE~\ref{table-csecs})
were obtained by combining the extracted Compton yields with the luminosity 
and detector acceptance. Figure~\ref{csec3} 
shows the total Compton scattering cross sections from the first and the second running 
period, respectively as a function of the beam energy. The extracted cross sections are compared to a next-to-leading order
calculation for both running periods.
All the results agree with the theoretical calculations within the experimental uncertainties.\\

The average total systematic uncertainty for each data point is 1.5\% for the first running period
and is 1.6~-~2.0\% for the second running period depending on the target (lowest for the 5\%
r.l. $^{12}$C target and highest for the 10\% r.l. $^{28}$Si target). The breakdown of the uncertainties is summarized
in Table~\ref{table-error}. The uncertainty in the photon flux is the largest source of 
uncertainty~\cite{Teymurazyan}. It was determined from the long term overall stability of the beam, data acquisition live time,
and tagger false count rate.
The uncertainty due 
to background subtraction was estimated from the variation in the fitting uncertainty with 
changes to the shape of the background distributions. The systematic uncertainty due to detector response was estimated from the change in experimental yield when the detector resolutions were varied by $\sim$3\%. The geometrical acceptance uncertainty was 
estimated from the variation in the simulated yields with small changes to the experimental 
geometry. The target thickness uncertainty was 0.05\% for the 5\% r.l. $^{12}$C 
target. The uncertainty was higher for the thicker targets used during the second
running period: 0.11\% for the 8\% r.l. $^{12}$C target and 0.35\% for 
the 10\% r.l. $^{28}$Si target~\cite{PrimEx74}. \\

The differences in systematic uncertainties for the two running periods stem from 
the differences in the experimental setup ( {\it{e.g.}} the geometry, the trigger ) and
differences in data analysis ( {\it{e.g.}} the energy binning, event selection, and the background fits ).

\section{V. Conclusion}

In conclusion, the total cross section 
for Compton scattering on $^{12}$C and $^{28}$Si, in the 4.400 - 5.475 GeV-energy range was
measured with the {\em PrimEx} experimental apparatus. The results are in excellent 
agreement with theoretical prediction with NLO radiative corrections. 
Averaged over all data points per target, the total uncertainties were 1.7\% for the first 
running  period, and 1.7\%, 2.0\%, and 2.6\% for the second running period (for 5\% r.l. and 
8\% r.l. ${}^{12}$C, and ${}^{28}$Si targets, respectively - see Table~\ref{table-error}).
This measurement provides an important verification of the magnitude and the sign of the radiative effects 
in the Compton scattering, which was determined and separated from the leading order process for the first 
time. We conclude that this measurement constitutes the first confirmation that the QED next-to-leading
order prediction correctly describes this fundamental process up to a photon energy, $E_{\gamma}$, 
of 5.5 GeV within our experimental precision.
\section{VII. Acknowledgements}
This work was funded in part by the U.S. Department of Energy, including contract %\# 
AC05-06OR23177 under which Jefferson Science Associates, LLC operates Thomas Jefferson 
National Accelerator Facility, and by the U.S. National Science Foundation (NSF MRI 
PHY-0079840). We wish to thank the staff of Jefferson Lab for their vital support 
throughout the experiment. We are also grateful to all granting agencies providing funding support to authors throughout this project.

\end{document}